\documentclass[fleqn,10pt]{wlscirep}
\usepackage[utf8]{inputenc}
\usepackage[T1]{fontenc}
\usepackage{subfigure}
\usepackage{graphicx}
\title{Heterogeneous model for superdiffusive movement of dense-core vesicles in {\it C. elegans}}

\author[1,2]{Anna Gavrilova}
\author[2]{Nickolay Korabel}
\author[1]{Victoria J. Allan}
\author[2,*]{Sergei Fedotov}
\affil[1]{School of Biological Sciences, Faculty of Biology, Medicine and Health, University of Manchester, The Michael Smith Building, Rumford St, Manchester, UK, M13 9PT. }
\affil[2]{Department of Mathematics, Faculty of Science and Engineering, The University of Manchester, Manchester, UK, M13 9PL.}
\affil[*]{Corresponding author: sergei.fedotov@manchester.ac.uk}

\keywords{Heterogeneous anomalous diffusion, beta-binomial distribution, dense-core vesicles}

\begin{abstract}
Transport of dense core vesicles (DCVs) in neurons is crucial for distributing molecules like neuropeptides and growth factors. We studied the experimental trajectories of dynein-driven directed movement of DCVs in the ALA neuron \textit{C. elegans} over a duration of up to 6 seconds. We analysed the DCV movement in three strains of \textit{C. elegans}: 1) with normal kinesin-1 function, 2) with reduced function in kinesin light chain 2 (KLC-2), and 3) a null mutation in kinesin light chain 1 (KLC-1). 
 We find that DCVs move superdiffusively with displacement variance $var(x) \sim t^2$ in all three strains with low reversal rates and frequent immobilization of DCVs. The distribution of DCV displacements fits a beta-binomial distribution with the mean and the variance following linear and quadratic growth patterns, respectively. We propose a simple heterogeneous random walk model to explain the observed superdiffusive retrograde transport behaviour of DCV movement. This model involves a random probability with the beta density for a DCV to resume its movement or remain in the same position. To validate our model further, we measure the first passage time for a DCV to reach a certain threshold for the first time. According to the model, the first passage time distribution should follow a beta-negative binomial distribution with the same parameters as the DCV displacement distributions. Our experimental data confirm this prediction.

\end{abstract}
\begin{document}

\flushbottom
\maketitle

\thispagestyle{empty}

\section*{Introduction}

Dense-core vesicle (DCV) transport is critical for regulating cellular and tissue homeostasis, particularly in neurons where DCVs must be transported along great distances \cite{guedes2019axonal,brady2017regulation}. This transport is mediated by motor proteins kinesins and dynein that promote the movement of cargos along microtubules \cite{bharat2017capture,wong2012neuropeptide}. Anterograde transport, which moves away from the cell body towards the axon tip, is carried out by kinesins, and retrograde transport, directed from the axon tip towards the cell body, is driven by dynein. 
DCVs carry diverse cagros such as neuropeptides, monoamines and neurotrophic factors. They are essential for the transport of various molecules needed in neuronal growth, signalling, learning and development transportation during movement or ageing \cite{hammarlund2008caps,gondre2012cellular,randi2023neural,ripoll2023neuropeptidergic,speese2007unc}. Originating in the neuron soma, DCVs undergo fast bidirectional axonal transport and release their content in response to specific stimuli, with fusion occurring at synaptic and non-synaptic sites \cite{hammarlund2008caps,bharat2017capture,schlager2010pericentrosomal,shakiryanova2006activity,wong2012neuropeptide}.

\begin{figure}[t!]
\centering
\includegraphics[width=0.99
\textwidth]{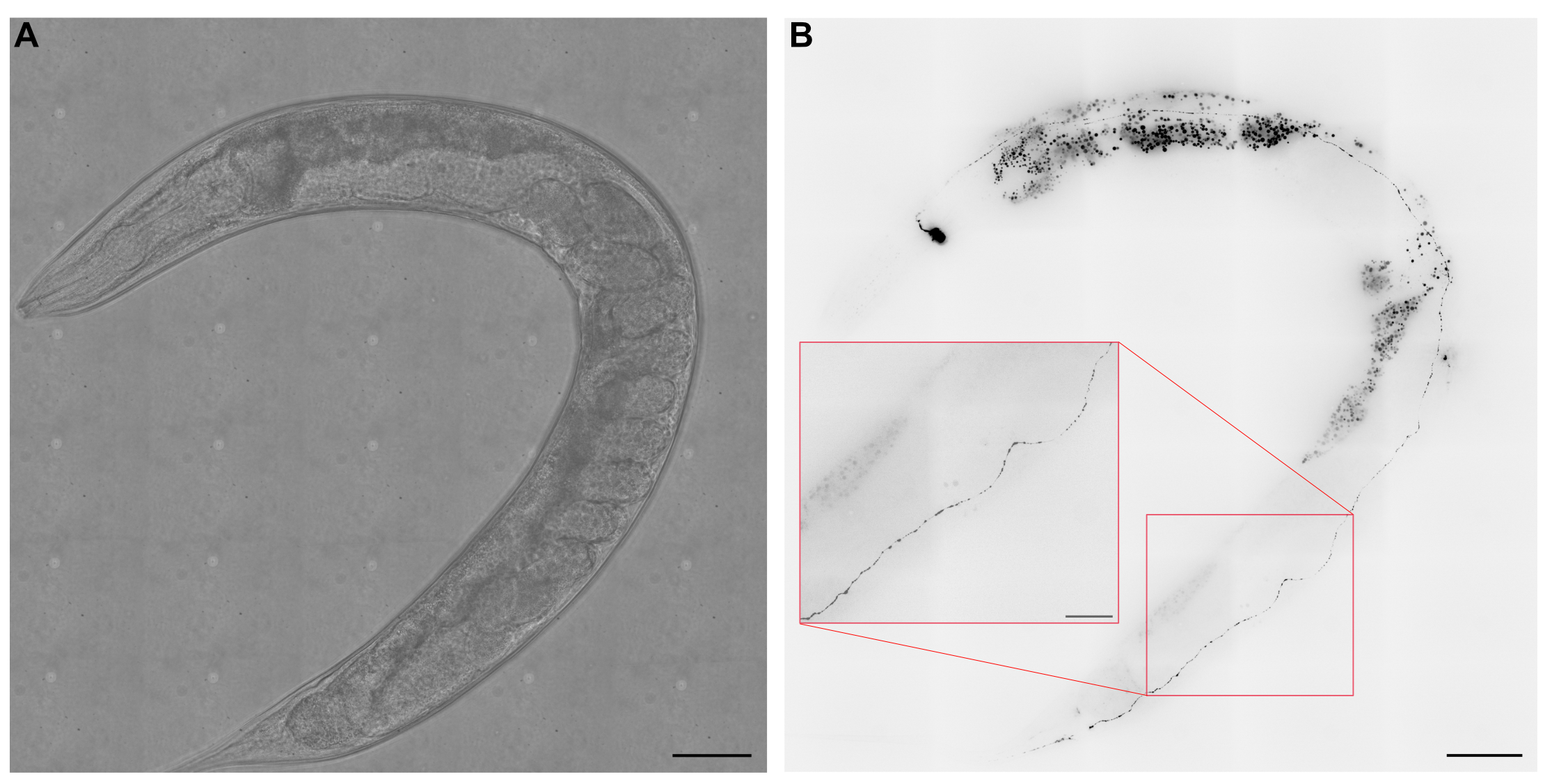}
\caption[Worm morphology and dense core vesicle (DCV) distribution]{Spinning disk brightfield {\bf A)} and fluorescence images {\bf B)} of an \textit{ida-1::gfp} worm showing the GFP-labelled  DCVs (inverted contrast). DCVs are shown as the black dots along the ALA neuron. An enlarged section of the ALA neuron is shown in the inset. Scale bars are 50 µm (20 µm in the inset).  The lighter spots in the brightfield images are artefacts generated by the tiling of multiple images. The larger particles visible outside the ALA neuron are auto-fluorescent granules, mainly located in the gut.}
\label{fig:zero}
\end{figure}

Here, we examine DCV mobility in the ALA neuron of nematode worms \textit{C. elegans}. The bidirectional nature of DCV mobility in the ALA neuron also agrees with data from other organisms \cite{gumy2017map2,schlager2010pericentrosomal,zahn2004dense}. The ALA neuron exhibits simple morphology, lacking dendrites typically present in multidendritic neurons, and has one of the longest neurites in the \textit{C. elegans} nervous system \cite{goodman2019caenorhabditis}. 
Given the small diameter of axons, moving DCVs occupy a considerable proportion of the total width \cite{ramirez2019axon}. 
Due to the narrow width of axons, DCVs occupy a significant portion of the axonal space, with stationary DCV accumulations observed along the axon in wild-type neurons where some DCVs move, pause, or stop \cite{bharat2017capture,hammarlund2008caps,schlager2010pericentrosomal,shakiryanova2006activity,wong2012neuropeptide}. 
The \textit{ida-1} gene in \textit{C. elegans}, an orthologue of mammalian type-1 diabetes auto-antigen proteins IA-2 and phogrin (IA-2$\beta$), is crucial in neuroendocrine tissues for insulin secretion and glucose metabolism \cite{zahn2001ida}. 
Transgenic \textit{ida-1::gfp} nematodes have previously been created by injecting N2 hermaphrodites with an \textit{ida-1p::ida-1::gfp} DNA plasmid construct \cite{zahn2001ida}. IDA-1 tagged with GFP is expressed in approximately 30 sensory neurons in \textit{C. elegans}, including ALA, and is a component of neuropeptide-containing DCVs \cite{zahn2001ida}. Confocal images of \textit{ida-1::gfp} with GFP-tagged DCVs in the ALA neuron are shown in Figure \ref{fig:zero}.
In the ALA neuron, DCVs exhibit predominantly unidirectional movement either towards or away from the cell body, with few reversals occurring within the axon \cite{gavrilova2024role}.

Here, we focus on dynein-driven DCV movement in three strains of \textit{C. elegans}: 1) with normal kinesin-1 function (called \textit{ida-1::gfp}), 2) with reduced function in kinesin light chain 2 (denoted as \textit{klc-2(rf)} or KLC-2), and 3) a null mutation in kinesin light chain 1 (denoted as \textit{klc-1(-)} or KLC-1). 
We use partial loss of function in KLC-2 because the null mutation is lethal.
Even though the mutations are in kinesin-1, it has been shown that the inhibition of kinesin-1 commonly reduces the velocity and frequency of movement in both directions \cite{gumy2017map2,hoerndli2013kinesin,gavrilova2024role}, meaning that dynein movement is also different compared to non-mutated worms as seen for DCVs in the ALA neuron. 
We chose to study dynein movement as KLC-1 and KLC-2 mutations have opposite effects on dynein movement, which makes the analysis more informative for modelling purposes.
We observe that the variance of DCV displacements grows proportionally to $t^2$. To describe this unusual superdiffusive behaviour, we develop a discrete random walk model with a stochastic probability of movement. Remarkably, this model explains the observed beta-binomial distribution of DCV displacements.
To further validate our model, we consider another key observable: the first passage time (FPT), which measures the time taken for a DCV to reach a defined threshold for the first time. Our model predicts that the FPT distributions (FPTDs) should follow the negative beta-binomial distribution with the same parameters as the distributions of DCV displacements, a prediction we confirm through analysis.
 
Most mathematical models of intracellular transport are based on concepts from Brownian motion, such as advection-diffusion, exclusion processes, and Brownian ratchet models \cite{bressloff2013stochastic,appert2015intracellular,julicher1997modeling}.
However, the complexity of motor-driven motility often leads to anomalous diffusion rather than classical Brownian motion, though this is not universally observed under all conditions \cite{caspi2000enhanced,tabei2013intracellular,kulkarni2006intracellular,kulic2008role,bruno2009transition,chen2015memoryless,reverey2015superdiffusion,song2018neuronal,flores2011roles,salman2002microtubules,klein2014fluctuation,korabel2021local,waigh2023heterogeneous,Kervrann}. 
One of the factors contributing to this complexity is the “tug-of-war” between kinesin and dynein motors, which are oppositely directed and act on the same cargo \cite{hancock2014bidirectional}. However, these dynamics are not yet fully understood. 
Furthermore, intracellular transport exhibits heterogeneity in both time and space, leading to varying non-Brownian behaviours that complicate modelling efforts \cite{han2020deciphering,korabel2021local,waigh2023heterogeneous}.
To address this, mathematical models of anomalous intracellular transport have been developed, including non-Markovian random walk models \cite{korabel2023non}, ensemble heterogeneity of diffusion coefficients \cite{korabel2023ensemble}, space-dependent diffusivity \cite{cherstvy2014particle}, and heterogeneity of jump probability \cite{fedotov2023population}. 
Additionally, a compartment model for DCV transport in type II axonal terminals of {\it Drosophila} motoneurons on a large time scale has been developed \cite{kuznetsov2017dense}. 
Finally, machine learning methods are currently being developed to analyse anomalous diffusion in experimental datasets  \cite{han2020deciphering,munoz2021objective,quiblier2024enhancingfluorescencecorrelationspectroscopy}.

In our recent paper \cite{gavrilova2024role}, we studied the effect of kinesin-1 mutations on the transport of DCVs in ALA neurons in \textit{C. elegans}. We showed how kinesin-1 mutations affect the lifespan and locomotion abilities of mutant worms. Here we focus on dynein-driven minus-end directed movement. The aim of this paper is to provide a simple data-driven model that offers insights into the heterogeneous movement of DCVs on a short time scale.

\section*{Results}

\begin{figure}[h!]
\centering
\includegraphics[width=1\textwidth]{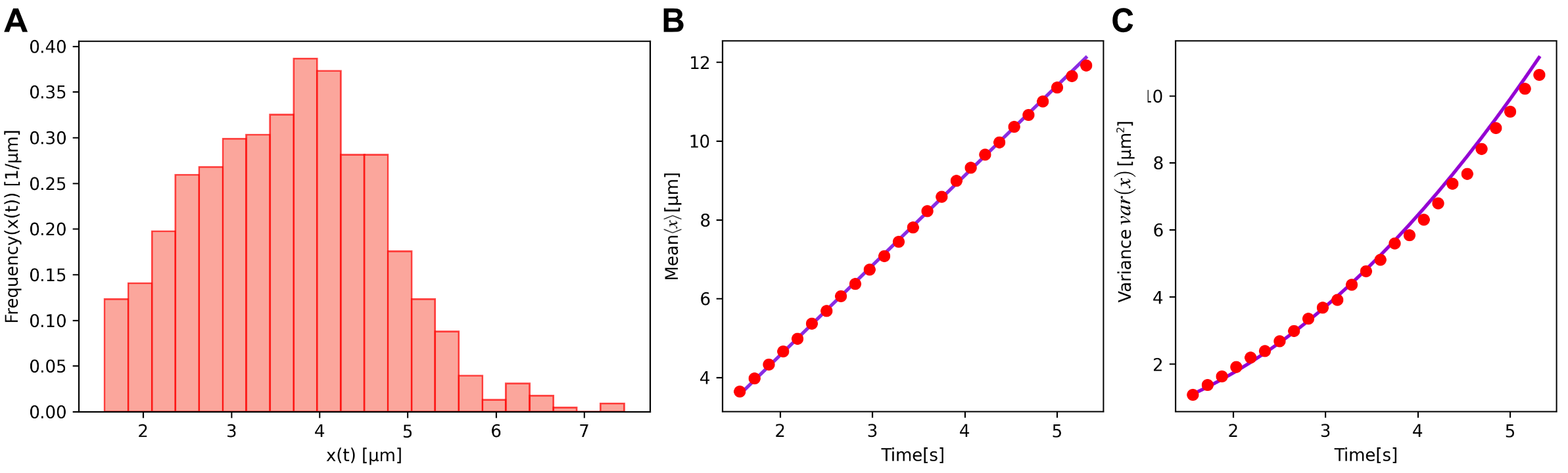}
\caption[Distribution of displacements $x(t)$ of DCV.]{\textbf{A} Distribution of displacements $x(t)$ of dynein-driven DCV in the ALA neuron of \textit{ida-1::gfp} worms calculated at time $t=1.563$ s. Mean $\left<x\right>$ \textbf{B} and variance $var(x)$ \textbf{C} of DCV displacements in the ALA neuron of\textit{ida-1::gfp} as the function of time. Symbols in \textbf{B} and \textbf{C} correspond to experimental data. Solid lines in \textbf{B} and \textbf{C} represent linear and quadratic functions, respectively, which follow equation (\ref{MeanVar}) as explained later in the text. Red dots are data points.} 
\label{fig:example_ida1}
\end{figure}

Our goal was to determine the empirical distribution of displacements 
$x(t)$ of dynein-driven DCVs and analyse its statistical characteristics such as the empirical mean and variance. Another aim was to develop a heterogeneous random walk model to potentially describe these data. We find that the distribution of $x(t)$ (Figure \ref{fig:example_ida1}A) has a unimodal form with high dispersion, indicating heterogeneity of the populations of DCVs. While the empirical mean shows linear growth in time as expected for a random walk, the variance unexpectedly shows a superdiffusive increase with time (Figure \ref{fig:example_ida1}B, C). This behaviour drastically deviates from the expected linear increase of variance. This suggests that the movement of DCVs cannot be accurately described by a standard random walk model. As the variance of displacements of retrograde tracks resembles a quadratic behaviour, this indicates that the DCVs perform superdiffusive ballistic movement.

We observed that DCVs predominantly move unidirectionally with frequent immobilization of DCVs. To describe this random movement, we used a simple discrete random walk model. The DCV starts at zero, takes a small step $a$ in a short time interval $\tau$ on the right with the probability $p$ (successful movement), or remains at the same position with the probability $1-p$ (failed movement). Then the probability mass function for the position of the DCV $x=ka$ at the time $t=n \tau$ is given by the binomial distribution $ P(x,t|p) = \binom{\frac{t}{\tau}}{\frac{x}{a}}p^{\frac{x}{a}}(1-p)^{\frac{t}{\tau}-\frac{x}{a}}.$
To model the heterogeneity of the DCV population, we assume that the probability $p$ is a random parameter with the beta-density $f(p)=B^{-1}\,\,(\alpha,\beta)p^{\alpha-1}(1-p)^{\beta-1}$ 
with $\alpha,\beta > 0$, where $B$ is the beta function.
Then one can obtain the well-known beta-binomial distribution\cite{skellam1948probability,johnson2005univariate}
\begin{equation}
  \bar{P}(x,t) = \int_0^1 P(x,t|p) f(p)dp = \binom{\frac{t}{\tau}}{\frac{x}{a}} \frac{B(\frac{x}{a}+\alpha,\frac{t}{\tau}-\frac{x}{a}+\beta)}{B(\alpha,\beta)},
  \label{BB}
\end{equation}
where $t=0,\tau, 2\tau,...$ and $x=0, a, 2a,...$.
The mean, $\left<x\right>$ and the variance, $var(x)$ are
\begin{equation}  \left<x\right>=\frac{\alpha}{\alpha+\beta}\frac{at}{\tau},  \quad var(x)=\frac{\alpha \beta}{(\alpha +\beta)(1+\alpha +\beta)} \frac{a^2 t}{\tau}+ \frac{\alpha \beta}{(\alpha +\beta)^2 (1+\alpha +\beta)} \frac{a^2 t^2}{\tau^2}.
\label{MeanVar}
\end{equation}

The essential feature of our heterogeneous DCV transport model is that the probability of successful movement $p$ is random. Assuming the randomness of this parameter, we use an indirect method to account for various factors influencing the movement of DCVs, including the size of DCVs, the level of activity of motor proteins, the structure of microtubules, etc. We assume that $p$ follows a beta distribution, with parameters $\alpha$ and $\beta$ determining the shape of the distribution.
The parameter $\alpha$ quantifies the level of successful movement of DCVs. A higher value of $\alpha$ means higher expected mobility of DCVs. The parameter $\beta$ measures the degree of failure in DCV movement. A higher value of $\beta$ corresponds to a situation where DCVs are immobilized, leading to a beta distribution skewed towards lower probabilities.
By adjusting the values of $\alpha$ and $\beta$, one can generate a wide variety of beta distributions for effective modelling of the heterogeneity of DCV movement. This fits the general idea of the heterogeneity in biological particle movement, which refers to the variability observed in individual particle behaviour \cite{waigh2023heterogeneous,fedotov2023population}. It signifies that all DCVs exhibit different movement patterns or characteristics.

\begin{figure}[ht]
\centering
\includegraphics[width=\linewidth]{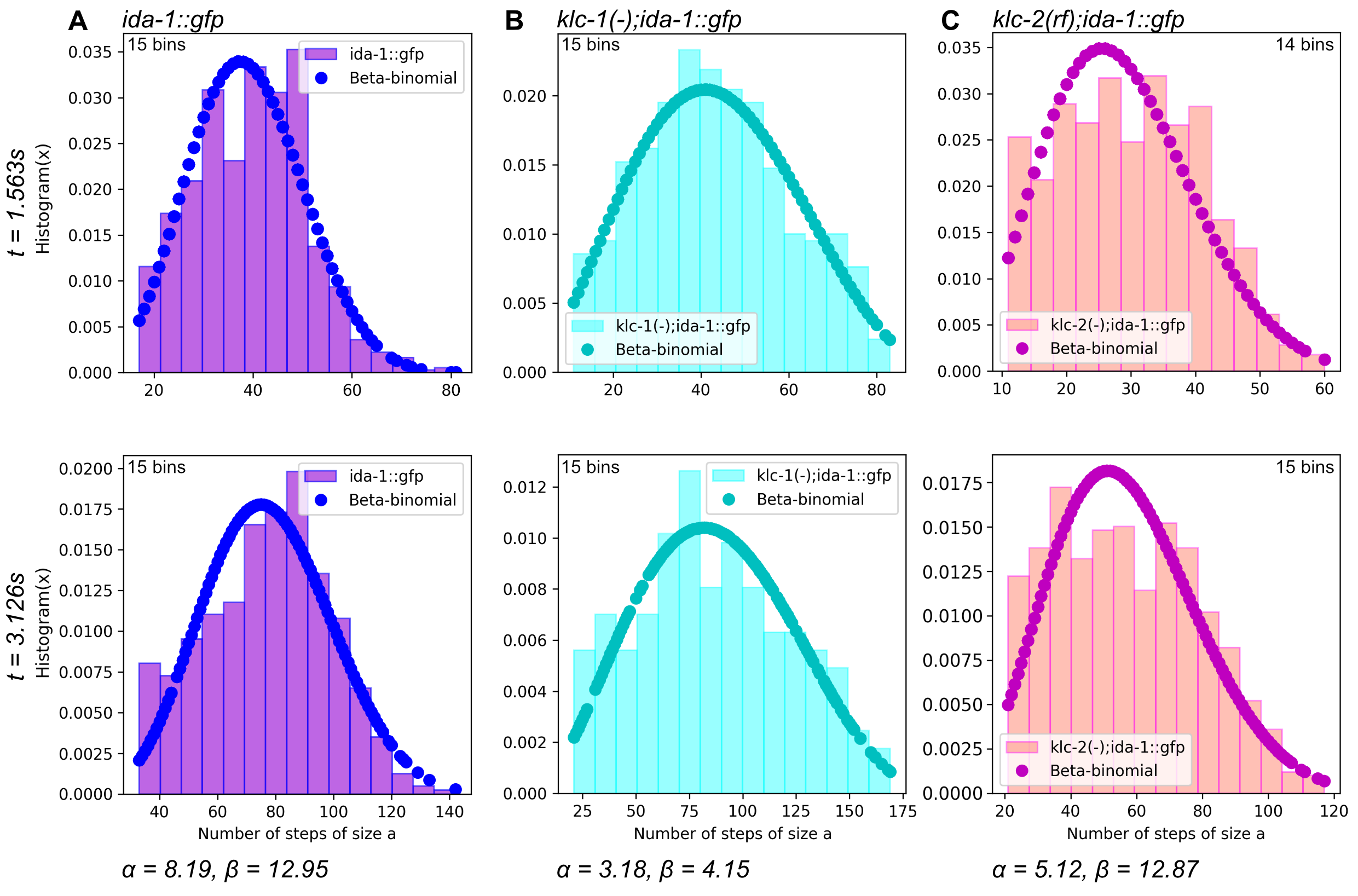}
\caption[Histograms of displacements of DCV movement with beta-binomial distribution fit ]{
Histograms of displacements of DCV movement with beta-binomial distribution fit. This figure shows histograms representing the dynein-driven displacement of DCVs for  \textbf{A} \textit{ida-1::gfp}, \textbf{B} \textit{klc-1(-)}; \textit{ida-1::gfp}, and \textbf{C} \textit{klc-2(rf)}; \textit{ida-1::gfp} strains measured at $t=1.563$ s and $t=3.126$ s. The number of tracks used to generate histograms for 1.563 s and 3.126 s: \textit{ida-1::gfp}- 851 and 549, \textit{klc-1(-)}; \textit{ida-1::gfp}- 438 and 289, and \textit{klc-2(rf)}; \textit{ida-1::gfp}- 1118 and 780 tracks respectively. 
The bars indicate the frequency distribution of DCV displacements, measured in the number of steps of pixel size $a=0.092$ $\mu$m.
The dots overlaid on the histograms represent the beta-binomial distribution fitted to the data with $\tau=0.01563$ s. The fitting parameters are for \textbf{A} $\alpha =8.19, \, \beta = 12.95$, \textbf{B} $\alpha =3.18, \, \beta = 4.15$,and \textbf{C} $\alpha =5.12, \, \beta = 12.87$.}
 
\label{fig:fourth}
\end{figure}

We fitted the beta-binomial distribution (\ref{BB}) to the empirical DCV dynein-driven displacement distribution by finding the optimal parameters $\alpha$ and $\beta$
using Python. This was done through maximum likelihood estimation.
The empirical displacement distribution for $t= 1.563 s$ and $t=3.126 s$  and fitted beta-binomial distribution are shown in Figure \ref{fig:fourth}.
The resolution of the camera used for imaging is 1 pixel = 0.092 $\mu$m, with a frame rate of 0.1563 seconds. Consequently, the time $\tau$ and the step size $a$ in (\ref{BB}) are chosen as $\tau=0.01563 $ s and $a=0.092$ $\mu$m.  
The number of trials $n$ in the standard beta-binomial distribution $n=t/\tau$, so at the time $t=1.56$ s in our random walk model DCV makes $n=100$ attempts to move to the right with the step size  $a=0.092$ $\mu$m. The number of tracks used to generate histograms (see Fig. \ref{fig:fourth}) for 1.563 s and 3.126 s is as follows:  \textit{ida-1::gfp}- 851 and 549, \textit{klc-1(-)}; \textit{ida-1::gfp}- 438 and 289, and \textit{klc-2(rf)}; \textit{ida-1::gfp}- 1118 and 780  tracks respectively. Since the sample sizes used to generate histograms are relatively small, the heights of neighbouring bins exhibit high variability.  
The fitting parameters of beta-binomial distributions were $\alpha =8.19, \, \beta = 12.95$ for \textit{ida-1::gfp}  (Fig. \ref{fig:fourth} panel A), $\alpha =3.18, \, \beta = 4.15$ for \textit{klc-1(-)}; \textit{ida-1::gfp} (panel B)  and $\alpha =5.12, \, \beta = 12.87$ for \textit{klc-2(rf)}; \textit{ida-1::gfp} (panel C). These parameters were determined by minimizing the negative sum of the log-likelihood function. 
Different values of $\alpha$ and $\beta$ reflect the different amounts of successful and failed movements of DCVs. 
The mean of successful movement, $\left<p\right> = \alpha/ \alpha+ \beta$, shows that the mean displacement in panel A is bigger than in panel C, as $\alpha$ in panel A is greater than in panel C, while the $\beta$ values are similar. In contrast, smaller $\beta$ in panel B results in increased variance, as follows from the equation \eqref{MeanVar}.

\begin{figure}[ht]
\centering

\includegraphics[width=1\textwidth]{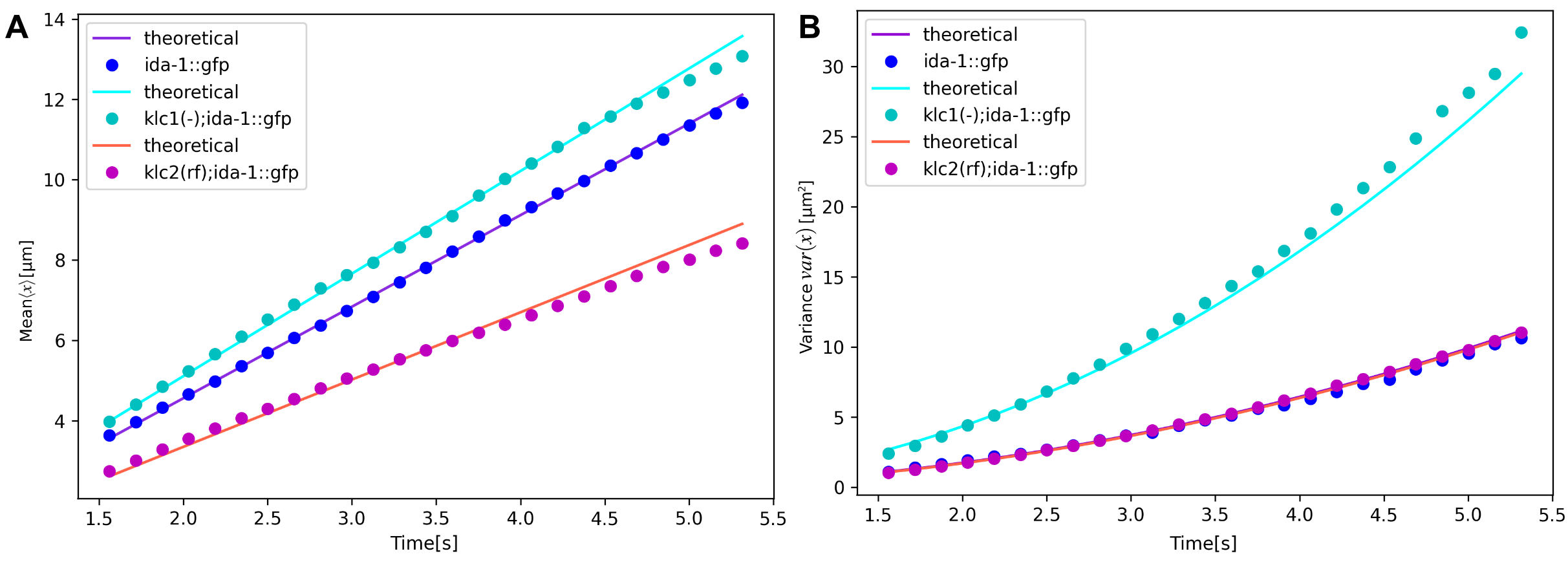}

\caption[Experimental and theoretical mean fitting]{Mean (\textbf{A}) and variance (\textbf{B}) of displacements of DCVs in the ALA neuron of \textit{ida-1::gfp} (blue dots), \textit{klc-1(-)}; \textit{ida-1::gfp} (cyan dots), and \textit{klc-2(rf)}; \textit{ida-1::gfp} (magenta dots) worms as function of time. 
Theoretical predictions for each data set (shown in lines) were calculated using equation \eqref{MeanVar} with parameters $\alpha$ and $\beta$ given in the text. }
\label{fig:second}
\end{figure}

Figure \ref{fig:second} illustrates the excellent agreement between the theoretical mean and variance (equation \eqref{MeanVar}) and the empirical displacement data for each strain. Surprisingly, a simple discrete random walk model accurately captures the empirical ballistic superdiffusive behaviour of the displacement variance over a wide time span, from 1.5 to 5.5 seconds. This reflects the remarkable robustness of the heterogeneous mechanism leading to the beta-binomial distribution. DCV movement in the wild type and KLC-2 strains show similar variability of displacements, although, in the KLC-2 strain, they move more slowly. The KLC-1 strain exhibits faster DCV movement but with greater variability.

\begin{figure}[ht]
\centering
\includegraphics[width=1
\textwidth]{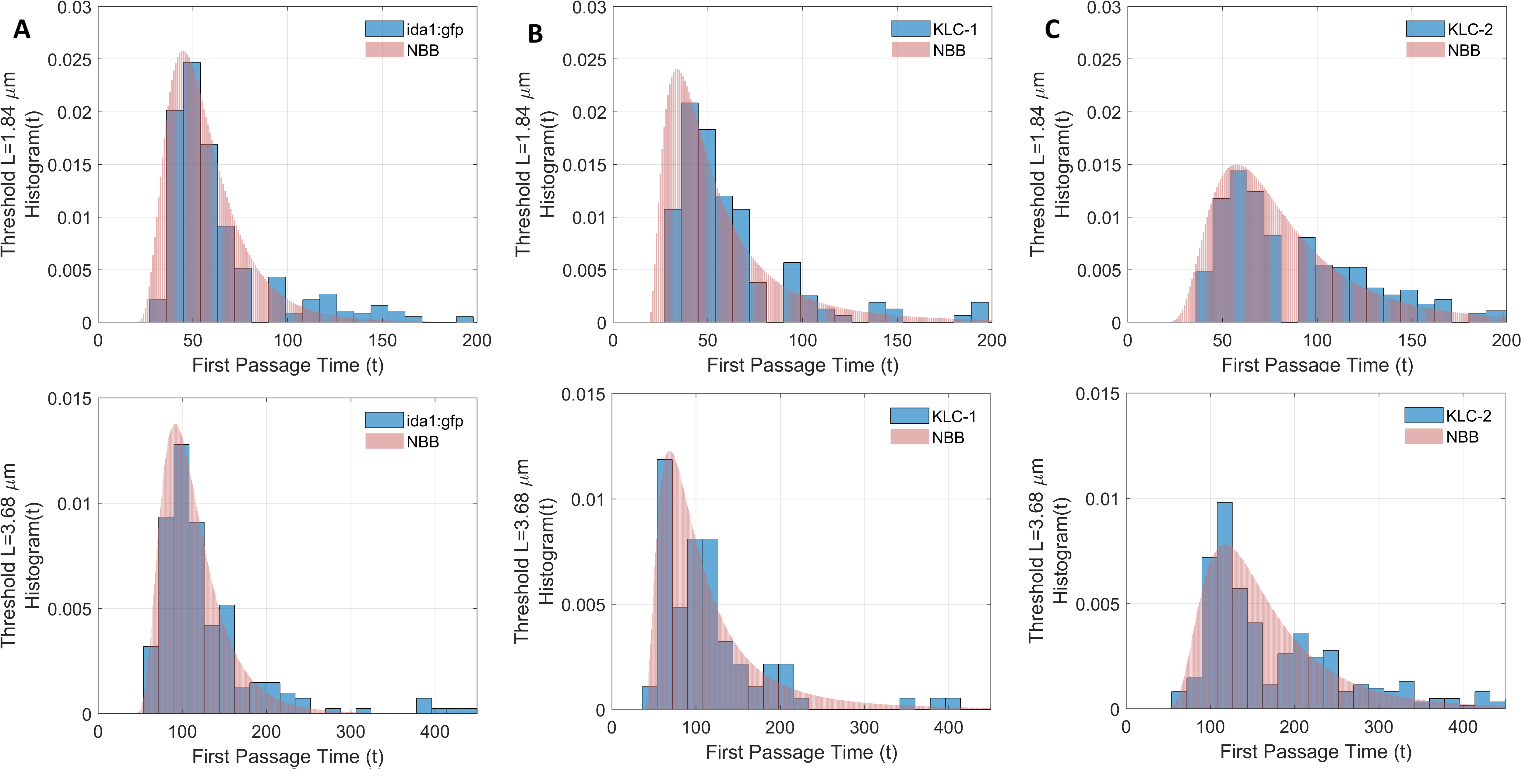}
\caption{
Histograms of first passage times $t$ to reach a threshold $L$ for the dynein-driven displacement of DCVs in   \textbf{A} \textit{ida-1::gfp}, \textbf{B} KLC-1 and \textbf{C} KLC-2 strains measured at fixed thresholds $L=1.84$ $\mu$m and $L=3.68$ $\mu$m. 
The overlaid on the histograms functions represent the beta-negative binomial distribution with the same parameters as in Figure \ref{fig:fourth}, namely, for \textbf{A} $\alpha =8.19$, $\beta = 12.95$, \textbf{B} $\alpha =3.18$, $\beta = 4.15$ and \textbf{C} $\alpha =5.12$, $\beta = 12.87$.}
\label{fig:third}
\end{figure}

\subsection*{First passage time distribution}

To test our model further, we consider another observable, namely, the first passage time (FPT) for a DCV to reach a threshold for the first time and consider their distributions. Let $F(t,z|p)$ be the conditional probability for the DCV first arriving at fixed position $z=ka$  at time $t=n \tau$, where $k$  is the number of steps for jumping right, $n\geq k$. 
%and denotes the total steps of the particle %performing.
Therefore,  $F(t,z|p)$ is the negative binomial distribution, that is, 
\begin{equation}
 F(t,z|q) =\binom{\frac{t}{\tau}-1}{
 \frac{z}{a}-1} q^{\frac{z}{a}} 
 (1-q)^{\frac{t}{\tau}-\frac{z}{a}}
 \label{conditionalFPT}
\end{equation}
Let $\overline{F}(t,z)$ be the average probability of $F(t,z|p)$ for $p$ obeying a beta distribution. Then
\begin{equation}
   \overline {F}(t,z) = \binom{\frac{t}{\tau}-1}{\frac{z}{a}-1} \int_0^1p^{\frac{z}{a}}(1-p)^{\frac{t}{\tau}-\frac{z}{a}}f(p)dp  = \binom{\frac{t}{\tau}-1}{\frac{z}{a}-1}\frac{B(\frac{z}{a}+\alpha, \frac{t}{\tau}-\frac{z}{a}+\beta)}{B(\alpha,\beta)} 
  \label{FPTD}
\end{equation}
for 
$t=\frac{\tau z}{a}, \frac{\tau z}{a}+\tau, \frac{\tau z}{a}+2 \tau,...$. This is a well-known beta-negative binomial (BNB) distribution \cite{johnson2005univariate}. 

One can find the mean $<t>$ of the first arrival time $t$
\begin{equation}
 <t>=  \frac{\tau z}{a} \, \frac{\alpha +\beta -1}{\alpha -1}  \quad \alpha > 1 
\end{equation}
and the variance 
\begin{equation}
 var(t)=\beta \frac{\tau^2 z}{a}\, \frac{(\frac{z}{a}+\alpha +1)(\alpha+\beta+1)}{(\alpha -2)(\alpha -1)^2}, \quad \alpha > 2.  
\end{equation}
For $\alpha \leq 1$ the mean value  $<t>$ does not exist. Thus, our model predicts that if the distributions of DCV displacements follow beta-binomial distributions, the FPT distributions should be described by the beta-negative binomial distributions with the same parameters $\alpha$ and $\beta$ as the distributions of DCV displacements.  

Figure  \ref{fig:third} shows good agreement between the empirical histograms of FPTs $t$ calculated for two threshold values $L=1.84$ $\mu m$ and $L=3.58$ $\mu m$ and the beta-negative binomial distributions (\ref{FPTD}) with the same parameters $\alpha$ and $\beta$ as DCV  displacement distributions. This confirms that our simple discrete random walk model for a heterogeneous population of DCV movement indeed describes the experimental data well.

\section*{Discussion}

In this study, we investigated the dynamics of the dynein-driven movement of dense core vesicles (DCVs) within the ALA neuron of three strains of \textit{C. elegans}: those with normal kinesin-1 function, reduced function in kinesin light chain 2 (KLC-2), and a null mutation in kinesin light chain 1 (KLC-1). 
Importantly, these mutations affect dynein function as well as kinesin-1 activity \cite{gavrilova2024role}. 
Our findings highlighted significant heterogeneity in DCV superdiffusive movement %\textcolor{green}{velocities} 
across all strains characterized by variances of displacements with quadratic growth in time ($var(x) \sim t^2$).  
Our results demonstrated that in all three strains, the traditional random walk models were insufficient to describe the observed movement patterns of DCVs. The quadratic growth in variance suggests that DCV transport mechanisms involve a heterogeneous population of DCVs with different mobility properties. This is particularly evident in the differing impacts of the mutations on DCV motility. In strains with normal kinesin-1 function, DCV movement exhibited relatively stable and consistent transport patterns, consistent with efficient motor function. However, in the KLC-1 null mutation strain, we observed increased variability and a broader distribution of displacements. The KLC-2 reduced function strain showed the opposite effects, with slower DCV movement but variance similar to the wild-type worms (Fig. \ref{fig:second}). These observations highlight the differential roles of kinesin light chains in modulating motor protein interactions and cargo transport efficacy in both directions.

The mobility of DCVs is influenced by many factors, such as the absence or changed activity level of motor proteins, the size and nature of the cargo carried by DCV vesicles, various proteins responsible for regulating DCV transport, etc. Indirect addressing of these important mechanisms within a single model can be accomplished using adjustments of parameters of the beta distribution. If we choose $\alpha > \beta$, the beta distribution is right-skewed, describing the situation of high mobility of DCVs. If $\alpha < \beta,$ the beta distribution is left-skewed, which describes more immobilization of DCVs. For example, when $\alpha  \to 0$, it means complete immobilization.
Small values of parameters $\alpha$ and $\beta $ lead to a highly dispersed distribution describing high heterogeneity in DCV mobility. 
As an illustration, different sizes and natures of the cargo carried by DCVs can be taken into account in the beta-binomial distribution.

A beta-binomial distribution (\ref{BB}) gave an excellent fit in the case of DCV displacement data for all strains. 
This model more accurately captures the probabilistic nature of motor-driven transport compared to conventional models, as it explicitly incorporates the inherent variability (heterogeneity) across different transport events within the population. It also accounts for the discrete nature of transport, where individual movements occur in distinct, quantifiable steps rather than as continuous processes. 
For each strain, we identified the best-fit parameters $\alpha$ and $\beta$, which determine transport patterns and provide insight into underlying molecular mechanisms using optimisation. 
Our results have broader significance for the mechanistic understanding of heterogeneous intracellular transport.
The superdiffusive quadratic growth in the variance of DCV displacements and the suitability of the beta-binomial distribution (\ref{BB}) underscore the complexity of motor protein dynamics and their regulation in a more realistic yet simple statistical model. 

%{\color{green}
%In this study, we developed and validated a heterogeneous model of dynein-driven DCV transport, incorporating the first passage time distribution as a key observable to characterize the transport dynamics. This finding offers valuable insights into the probabilistic nature of motor-driven transport, a key feature of intracellular transport systems.}

%green
A beta-negative binomial distribution also fits the experimental FPT distributions, as predicted by the model, providing strong support for the validity of the model. FPTs play a crucial role in regulating the timing of DCV delivery to synaptic terminals and other target sites within neurons, where they transport neuropeptides and growth factors vital for synaptic modulation, axonal growth, and neuronal survival. Therefore, the precise timing of these deliveries, as reflected by FPTs, is essential for maintaining proper neuronal function, and disruptions in transport timing can lead to impaired synaptic activity, hindered axonal growth, and contribute to neurodegenerative diseases such as Alzheimer's and Parkinson's \cite{brady2017regulation,sleigh2017methodological}. %Our findings highlight the importance of optimizing transport dynamics, including minimizing FPTs, to ensure efficient cargo delivery, which is necessary for maintaining cellular homeostasis and adapting to changes in neuronal activity. 
%Thus, therapeutic strategies that target motor protein activity or improve cytoskeletal organization could help restore proper transport timing, reduce variability in FPTs, and enhance transport efficiency, offering new avenues for treating diseases characterized by compromised intracellular transport.

Our model captures the inherent heterogeneity in transport dynamics, as evidenced by the substantial variation in FPTs observed across the DCV population. This variability aligns with the known heterogeneity in motor-driven movement, which arises from factors such as differential motor protein activity, interactions with the cytoskeleton, and variations in the local cellular environment. By explicitly accounting for this population-level variability, our model offers a significant improvement over traditional models that typically assume uniform transport behaviour. This enhanced approach provides a more accurate representation of the complexity of motor-driven transport, facilitating a deeper understanding of the factors contributing to variability in DCV movement and its impact on overall transport efficiency.

In conclusion, this study provides a detailed framework for understanding the heterogeneous dynamics of motor-driven DCV transport, offering new insights into the regulation of neuronal activity and transport efficiency. By quantifying the variability in transport events, our model allows for a more nuanced understanding of intracellular transport mechanisms and their implications for neuronal health and disease. Future work will further refine this model and explore its application in different cellular contexts, with the ultimate goal of uncovering new therapeutic strategies for diseases associated with impaired intracellular transport.

\section*{Methods}

\subsection*{\textit{C. elegans} strains}

As a model organism, we use nematode worms {\it Caenorhabditis elegans,  (C. elegans)}.
We used a strain expressing IDA-1::GFP, which allows for robust GFP (green fluorescent protein) labelling of DCVs in the ALA neuron \cite{zahn2004dense} (Fig. \ref{fig:zero}). Kinesin-1 is a heterotetrameric molecular motor complex composed of two kinesin heavy chains (KHCs/KIF5) and two kinesin light chains (KLCs) \cite{anton2021molecular}. 
To compare the motility of DCV in \textit{C. elegans} with kinesin-1 mutations, we crossed the \textit{ida-1::gfp} strain with mutants displaying reduced function of kinesin light chain 2 \textit{(klc-2(km11)}, referred to here as \textit{klc-2(rf)}) \cite{sakamoto2005caenorhabditis}, and a kinesin light chain 1 deletion mutant \textit{(klc-1(ok2609)}, referred to as \textit{klc-1(-)}) \cite{c2012large}.
In the genome of \textit{C. elegans}, the \textit{unc-116} gene encodes a single kinesin heavy chain (KHC), and the \textit{klc-1} and \textit{klc-2} genes encode two kinesin light chains (KLCs) \cite{sakamoto2005caenorhabditis}.
\textit{C. elegans} strains were cultured and maintained according to standard protocols \cite{brenner1973genetics}. Worms were raised at 20°C on 6 cm plates (Greiner Bio-One) containing nematode growth media (NGM) seeded with OP-50 bacteria as food.
The \textit{klc-1(ok2609)} and \textit{klc-2(km11)} strains were obtained from the Caenorhabditis Genetics Centre (CGC), while the \textit{ida-1::gfp} strain was obtained from John Hutton’s laboratory at the University of Colorado Health Sciences Center, Denver, and kindly provided by Howard Davidson. Crosses of \textit{ida-1::gfp} with \textit{klc-1(ok2609)} and \textit{klc-2(km11)} were performed as described  \cite{gavrilova2024role}.

\subsection*{Imaging}

To observe DCV transport events in worms expressing IDA-1::GFP, we used time-lapse microscopy. Confocal imaging was conducted using a 3i inverted Spinning Disk Confocal Microscope equipped with a motorized stage for live-cell imaging. Images were captured with a CSU-X1 spinning disc confocal system (Yokogawa) on a Zeiss Axio-Observer Z1 microscope using various objectives, Prime 95B Scientific CMOS (1200 x1200 11 $\mu$m pixels;
backlit; 16-bit) camera (Photometrics) and motorised XYZ stage (ASI). Slidebook software (3i) was used for image acquisition.

Whole-worm images (Figure \ref{fig:zero}) were generated using Multiple XY Location Capture in SlideBook to produce montage images. Brightfield and fluorescence images were captured with the same camera, with LED brightfield illumination and a 100 ms exposure time. Fluorescence images were composed of multiple z-planes combined by maximum projection.

Videos of DCV movement were recorded using a 100×/1.30 Plan-Neofluar objective, with a single plane and a 156.3 ms frame interval for 500 frames. In each strain, videos were taken along the whole length of the ALA neuron as it was shown that distance from some does not affect the transport properties of DCVs \cite{gavrilova2024role}.
Raw TIFF files were stabilized using the Image Stabilizer plugin for ImageJ, with adjustments made as needed for significant worm movement. Kymographs were generated using the ImageJ Multi Kymograph plugin, and tracks were identified using KymoButler \cite{jakobs2019kymobutler}. The x and t positions for each track in every kymograph were obtained using the KymoButler code in Mathematica Version 14.0 (Wolfram Research, Inc.).

\subsection*{Analysis of the trajectories}

We combined KymoButler results for each strain and performed analysis and visualization using custom Python code (available at \url{https://github.com/umkich/organelle_transport_analysis}).  In instances where two x positions corresponded to the same t, indicating simultaneous particle occupancy, the average position was calculated. KymoButler outputs trajectories in pixels and timeframes format. We converted them to seconds and µm for further analysis (one timeframe corresponds to 0.1563 s and one pixel to 0.092 µm). 

To isolate dynein-driven retrograde tracks, we excluded stationary and kinesin-driven anterograde tracks. This was achieved by selecting a threshold near the local minimum between retrograde and stationary peaks for each strain. Since the movement characteristics varied between strains, the thresholds were different. For the \textit{ida-1::gfp} strain, we set the threshold at 1.6 $\times$ the number of frames. For the \textit{klc-1(-)};\textit{ida-1::gfp} and \textit{klc-2(rf)};\textit{ida-1::gfp} strains, we chose a threshold of 1 $\times$ the number of frames.

\bibliography{main}

\section*{Acknowledgements}

We extend our gratitude to Howard Davidson (University of Colorado Health Sciences Center, Denver) and the CGC for providing worm strains. The CGC is funded by the NIH Office of Research Infrastructure Programs (P40OD010440). A.G. would like to thank Gino Poulin (University of Manchester) for his assistance and guidance in working with \textit{C. elegans}. The 3i spinning disk microscope used in this study was purchased through the University of Manchester Strategic Fund. Special thanks are due to Dr. Peter March for his help with confocal imaging. We are also grateful to Dmytro Chekunov for his assistance with Python and Mathematica coding. This work was funded by a Wellcome Trust PhD studentship awarded to AG (Grant No. 108867/Z/15/Z). SF, NK and VA were supported by EPSRC (Grant No. EP/V008641/1).

\section*{Author contributions statement}

A.G. generated specific worm strains, imaged neurons with a spinning-disk confocal microscope, analysed DCV movement using KymoButler and custom Python code, segmented and calculated displacements, fitted a beta-binomial distribution to the data, prepared figures, and co-authoring the text of the paper. N.K. contributed to formal analysis, validation, investigation, and writing (review and editing). 
V.A. contributed to conceptualization, resources, data curation, supervision, funding acquisition, validation, methodology, project administration, and writing (review and editing). 
S.F. contributed to conceptualization, formal analysis, supervision, funding acquisition, validation, methodology, project administration, and writing (review and editing).

\section*{Additional information}

\textbf{Competing interests}: No competing interests declared.
\\
\textbf{Data availability}: The data supporting this study's findings are available upon request from the corresponding author.

\end{document}